\documentclass[english,aps,prl,amsmath,twocolumn,amssymb,nofootinbib,superscriptaddress,showpacs,floatfix,longbibliography]{revtex4-2}

\usepackage[english]{babel}

\usepackage[utf8]{inputenc}
\usepackage{latexsym}
\usepackage{braket}	
\usepackage{amsmath,amssymb,amsfonts,textcomp,latexsym,pb-diagram,amsopn,bm}
\usepackage{graphicx,xcolor}
\graphicspath{{pictures/}}
\DeclareGraphicsExtensions{.pdf,.png,.jpg,.eps}
\usepackage[breaklinks,colorlinks,bookmarks=false,citecolor=blue,linkcolor=red,urlcolor=blue]{hyperref}
\usepackage{url}
\usepackage{hyperref}
\usepackage{tabularx}
\newcolumntype{L}{>{\raggedright\arraybackslash}X}
\usepackage{epstopdf}

\begin{document}

\title{ Tunneling maps of interacting electrons in real time: \\ 
 anomalous tunneling in quantum point-contacts beyond the steady state regime}

\author{Taras Hutak, Gleb Skorobagatko*}

\affiliation{%
\mbox{%
Institute for Condensed Matter Physics of National Academy of Sciences of Ukraine,
  Svientsitskii Str.1,79011 Lviv, Ukraine%
  } \\
  *Correspondence to: gleb.a.skor@gmail.com}


\begin{abstract}
\begin{small}
Strongly correlated transport of interacting electrons through the one-dimensional tunnel contact is considered within the Luttinger liquid model of one-dimensional electrodes on arbitrary time scales $ t> \hbar /\Lambda_{g} $. Using previous results, namely, the Self-equilibration theorem proven in Ref.[\textit{G.~Skorobagatko,  Ann.Phys.,\textbf{422}, 168298 (2020)}] the exact distribution functions $ P(N,t) $ for the probabilities  of $ N $ electrons to tunnel through the contact during the fixed time interval $ t $ are derived and plotted as 3D intensity graphs called \textit{real-time tunneling maps} in a whole available region of Luttinger liquid correlation parameter values, bias voltages, and temperatures. The tunneling maps being obtained reveal \textit{tunneling anomalies}- unusually high tunneling probabilities for the small numbers of electrons on the short-time scales beyond the steady state regime of electron tunnel transport. Especially, this concerns the case of strong short-range electron-electron interaction in the leads at $ g \lesssim 1/2 $. The effect being found can be treated as the direct consequence of the \textit{self-equilibration} phenomenon. It is shown that features of such an anomalous electron tunneling beoynd the steady state regime can be used for precise experimental measurements of small differences between Luttinger liquid correlation parameter values in different one-dimensional quantum-point contacts.  
\end{small}
\end{abstract}

\maketitle
\begin{normalsize}


The problem of the effective quantum detection of qubit quantum states lies in the core of a contemporary progress in quantum information science and quantum computers design \cite{1,2,3,4,5,6}. Especially, this concerns charge-qubits constructed as spatial electron states being delocalized within two tunnel-connected metallic droplets which form a quantum double-dot in the regime of Coulomb blockade \cite{5,6,11,12,13}. In the latter case of our interest a biased quantum-point contact (QPC) plays the role of quantum detector \cite{6,11} of electron states on the double dot if one couples the former capacitively to a given charge qubit \cite{12,13}. In this case electron states on the double-dot are easy to prepare and to manipulate by external gates \cite{1,2,3,4} though the problem of decoherence due to back-action from quantum detector remains challenging in such type of charge-qubit design \cite{11,12,13}. 
Here the best situation for qubit states preparation and detection can be realized when one neglects the internal quantum dynamics of charge qubit due to very slow Rabi oscillations of the electron quantum state on the double dot with non-zero inter-dot tunneling in comparison with the characteristic time scales of qubit decoherence and acquisition of information by the QPC quantum detector \cite{6,11}. However, even in the experimental situations being that ideal one faces several fundamental problems which can considerably smear or even misrepresent the information about the qubit decoherence timescales and, hence, about the true qubit quantum states being detected by means of given QPC \cite{11,12,13}. In particular, one factor here is a well-known Anderson orthogonality catastrophe  due to Coulomb electrostatic interaction between the electron delocalized on the double dot and the electrons tunneling through the QPC quantum detector \cite{6,11}. This effect leads to a fast decoherence of any initially prepared charge-qubit state superposed from its states  $ \vert 1 \rangle $ and $ \vert 2 \rangle $  on the dot $ 1 $ and on dot $ 2 $, correspondingly, as compared to the characteristic timescales of information acquisition by the statistics of tunneling events for given QPC \cite{6,11}. Another basic effect for one-dimensional QPCs is Coulomb repulsion between the electrons in the one-dimensional geometry of quantum wires \cite{7,8,9}, which leads to a strong suppression of tunneling density of states at the edges of Luttinger liquid QPC electrodes in the close vicinity of the Fermi energy \cite{9,10}. This results in a well-known Kane-Fisher effect \cite{10,16} of tunnel current suppression at low temperatures, which simply nullifies tunnel current in the limit $ T=0 $ even at very small electron-electron repulsion in 1D leads. Remarkably, recently it has been shown \cite{11} that the interplay between these two effects 
can cause huge signal "delays" in QPC quantum detectors as compared to the true charge qubit decoherence time scales have been initially predicted. As well, several evidences of such "decoherence delays" has been observed experimentally \cite{12,13}, while their possible physical explanation has been proposed in Ref.~\cite{11}. Especially, in Ref.~\cite{11} this effect has been referred to as the consequence of the  \textit{quantum detection instability} which occurs in the effectively one-dimensional QPCs due to strongly correlated Luttinger liquid behaviour of tunneling electrons in very narrow "window" of Luttinger liquid correlation parameter  values $ 0.6 \lesssim g \lesssim 0.8 $ in the QPC electrodes (see Ref.~\cite{11} for details). Therefore, a precise experimental measurement of the Luttinger liquid correlation parameter $ 0< g \leq 1 $ in QPC quantum wires by means of various tunnel transport experiments \cite{14,15}  becomes crucial for the entire problem of charge-qubit quantum detection \cite{6,11,12,13}. However, the experimental estimations of  $ g $ in QPC quantum wires from tunnel transport measurements  of any kind remains a traditionally weak point due to considerable quantum fluctuations of tunnel current at low temperatures as well as due to the problem of averaging over the large periods of observation in order to extract $ g $ from different steady state characteristics of given QPC, such as e.g. average tunnel current at given bias voltage, shot-noise power, etc. \cite{14,15}. Moreover, it is known \cite{14} that a precise estimation of $ g $ is especially difficult for strongly interacting electrons in the 1D quantum wires with  $ 0< g \lesssim 1/2 $ (see e.g. Ref.~\cite{14}). On the other hand, the large time scales of the QPC steady state formation also prevent any fast enough quantum detection of charge qubits quantum states making problematic the effective external control over the entire charge qubit quantum dynamics. 

Therefore, here we propose the way in which one can solve both these problems providing: i) precise estimation of the small differences in the values of Luttinger liquid correlation parameter $ g $ for different tunnel contacts at low temperatures for the situation of strong enough electron-electron repulsion in these QPCs with $ 0< g \lesssim 1/2 $ and ii) fast enough quantum detection by means of QPC using the statistics of electron tunneling \cite{18,19}  on the short-time scales beyond the steady state regime \cite{20}. The key to these results is in the analysis of the probabilities $ P(N,t) $ for $ N $ electrons ($ N=1,2,3 \ldots $) to tunnel through given Luttinger liquid QPC during the fixed time interval $ t $ as functions of  all external parameters such as bias voltage $ V $ applied to QPC electrodes, Luttinger liquid correlation parameter $ g $  ($ 0< g \leq 1 $) and temperature. 
This scenario becomes possible due to validity of so-called \textit{S-theorem} \cite{11,20} for the exact calculation of Keldysh distribution function of interacting electrons which tunnel through a weakly linked QPC. 


Let us consider the quantum point-contact (QPC) between two semi-infinite one-dimensional quantum wires  in the weak tunneling limit \cite{6,11,16,20} 
\begin{equation}
H_{\Sigma}=H_{\text{LL}}+H_{\text{int}},
\label{eq1}
\end{equation}
where $H_{\text{LL}}$ is the Hamiltonian of the left ($L$) and right ($R$) Luttinger liquid quantum wires
and $H_{\text{int}}$ represents tunnel interaction between them \cite{16}. Without the loss of generality we assume that our QPC is located at the spatial point $x=0$ and each wire is connected to its own reservoir of non-interacting electrons at spatial point $ x=-L \rightarrow -\infty $. The difference between chemical potentials $ \mu_{L} $ and $ \mu_{R} $ of the left and right leads is proportional to the bias voltage $ V $ applied to the QPC $\mu_{L} - \mu_{R} = eV   $ \cite{16,17}.  Therefore, the Hamiltonian $H_{\text{LL}}$ is non-interacting in its bosonic representation \cite{9,10,16}
\begin{equation}
H_{\text{LL}}=\frac{1}{2\pi}\sum_{j=L,R}v_{g}\int\limits_{-\infty}^{0}dx
\bigg\{
g\Big(
\partial_{x}\varphi_{j}(x)
\Big)^{2}+
\frac{1}{g}
\Big(
\partial_{x}\theta_{j}(x)
\Big)^{2}
\bigg\},
\label{eq2}
\end{equation}
here as well as in all subsequent formulas $g$ is the dimensionless Luttinger liquid correlation parameter defined as $g=(1+U_{s}/2E_{F})^{-1/2}$ (where $g=1$ corresponds to Fermi liquid in the QPC leads at $U_{s}=0$, while $0<g<1$ is the case of strong Coulomb repulsion between electrons $U_{s}>E_{F}$, $ U $ is a  potential energy of short-range repulsion between electrons in the QPC leads) and  $v_{g}=v_{F}/g$ is the group velocity of collective plasmonic excitations in the leads (here $ v_{F} $ is the Fermi velocity (we put $ \hbar =1 $ everywhere for simplicity) \cite{9,11,16,20}. In Eq.~(\ref{eq2}) $\varphi_{L(R)}(x)=\pi\int_{-\infty}^{x}dx^{\prime}j_{L(R)}(x^{\prime})$ and $\theta_{L(R)}(x)=\pi\int_{-\infty}^{x}dx^{\prime}\rho_{L(R)}(x^{\prime})$ are common phase- and charge quantum bosonic fields with standard commutation relations $ [\theta_{\alpha}(x),\varphi_{\alpha'}(x')]=\frac{\text{i} \pi}{g} sgn(x-x') \delta_{\alpha,\alpha'} $ (here $ \alpha, \alpha'=L,R $) and zero average values with respect to a Luttinger liquid ground state $\langle\varphi_{L(R)}(x,t)\rangle=\langle\theta_{L(R)}(x,t)\rangle=0$ \cite{6,11,20}. 

In our model the presence of weakly linked tunnel junction in our 1D electron system at $ x=0 $ implies following boundary condition for bosonic charge- quantum fields at spatial point $x=0 $ for any moment of time \cite{16,17} 
\begin{equation}
\theta_{L}(x=0)=\theta_{R}(x=0)=0.
\label{eq3}
\end{equation}
Hence, our bosonized tunnel Hamiltonian, in the interaction representation in terms of spatially non-local bosonic field $ \varphi_{-}(x,t) = \varphi_{L}(x,t) - \varphi_{R}(x,t)$ takes following non-linear operator form \cite{6,11,20} (for more details one can see also Appendix A in the Supplementary material below) 
\begin{equation}
H_{\text{int}}(t)=\tilde{\lambda}\cos \left( \varphi_{-}(x,t) + eVt \right) \vert_{x=0},
\label{eq4}
\end{equation}
where $ \tilde{\lambda} = \vert t_{t} / (\pi a_{0})\vert$ - is the tunnel coupling constant for our bosonized QPC model (here $ t_{t} $ is a bare tunneling amplitude and $ a_{0} $ is the lattice constant in the leads). Weak tunneling limit means that in all subsequent formulas one has $ \tilde{\lambda}/ \Lambda_{g} \ll 1 $, where $ \Lambda_{g}=\Lambda/g \simeq E_{F}/g $ with $ E_{F} $ being the Fermi energy of electrons in both QPC leads. The bosonized tunnel Hamiltonian of the QPC in the limit of weak tunneling will account for all inter-plasmon interactions in our bosonized model in what follows.

Now in order to obtain time-dependent distribution functions $ P(N,t) $ for $ N $ electrons ($ N=1,2,.. $) to tunnel through a weakly linked QPC model of Eqs.~(\ref{eq1}-\ref{eq4}) during the fixed time interval $ t > 0 $ one needs to apply a well-known full counting statistics (FCS-) method \cite{18,19,20} to the bosonized QPC model of Eqs.~(\ref{eq1}-\ref{eq4}) in the interaction representation. The basic ingredient here is a so-called Keldysh distribution function (KDF) which is a sum  $ \tilde{\chi}(t,\xi)=\sum_{N=1}^{\infty} P(N,t)e^{\text{i} \xi N} $ over all possible tunneling realizations by the time $ t $ passed from the beginning of the observation \cite{19,20}. Here a variable $ \xi\in [-\pi;\pi] $ is electron's counting field - physically, it is an additional (magnetic) phase acquired by each elementary charge, while it tunnels through a given QPC \cite{19}. Alternatively, one can calculate KDF as following thermal average over the ground state of QPC leads \cite{6,20}  $ \tilde{\chi}(t) = \left\langle \mathcal{T}_{K} \exp\left( -\text{i}\int_{C_{K}(t)} H_{int}(\tau; \xi(\tau))d\tau\right)  \right\rangle  $,  where the symbol $ \mathcal{T}_{K}  $ denotes all possible arrangements of all time-dependent operators from the subsequent exponent along the time-dependent Keldysh contour $ C_{K}(t) $ in the complex time $ \tau $- plane \cite{6,11,20}. In our case of bosonized weakly linked Luttinger liquid junction KDF $ \tilde{\chi}(t,\xi) $ is defined by tunnel Hamiltonian (\ref{eq4}) in the interaction representation being modified by a counting field $ \xi $ as follows \cite{20} $ H_{\text{int}}(\tau; \xi(\tau)) = \tilde{\lambda}\cos \left( \varphi_{-}(0,\tau) + eV\tau + \xi(\tau)/2 \right)$ (see also Ref.~\cite{22}), where $ \xi(\tau) $ is equal either to $ +\xi $ or to $ -\xi $ dependent on the upper (or lower) branch of the Keldysh contour in the complex plane \cite{20}. 

Remarkably, the Self-equilibration (or SE-) theorem proven for the first time in Ref. \cite{20} states that the following re-exponentiation equality 
\begin{equation}
\left\langle \mathcal{T}_{K} \exp\left( -\text{i}\int_{C_{K}(t)} H_{int}(\tau; \xi(\tau))d\tau\right)  \right\rangle = \exp \left\lbrace  -\mathcal{F}(t,\xi) \right\rbrace
\label{eq5}     
\end{equation}
\textit{is exact} for any weakly linked Luttinger liquid QPC at arbitrary timescales $ t > 1/\Lambda_{g} $ (for details of the proof one can see Appendix B of Ref. \cite{20}), thus, giving rise to a non-trivial physical effect of \textit{self-equilibration} for quantum fluctuations in tunnel transport beyond the steady state regime \cite{20}. As the result, the exact formula for the cumulant- generating functional $ \mathcal{F}(t,\xi) $ from Eq.~(\ref{eq5}) has been calculated in Ref. \cite{20} \textit{exactly}. 

\begin{figure*}
\includegraphics[width=0.73\textwidth]{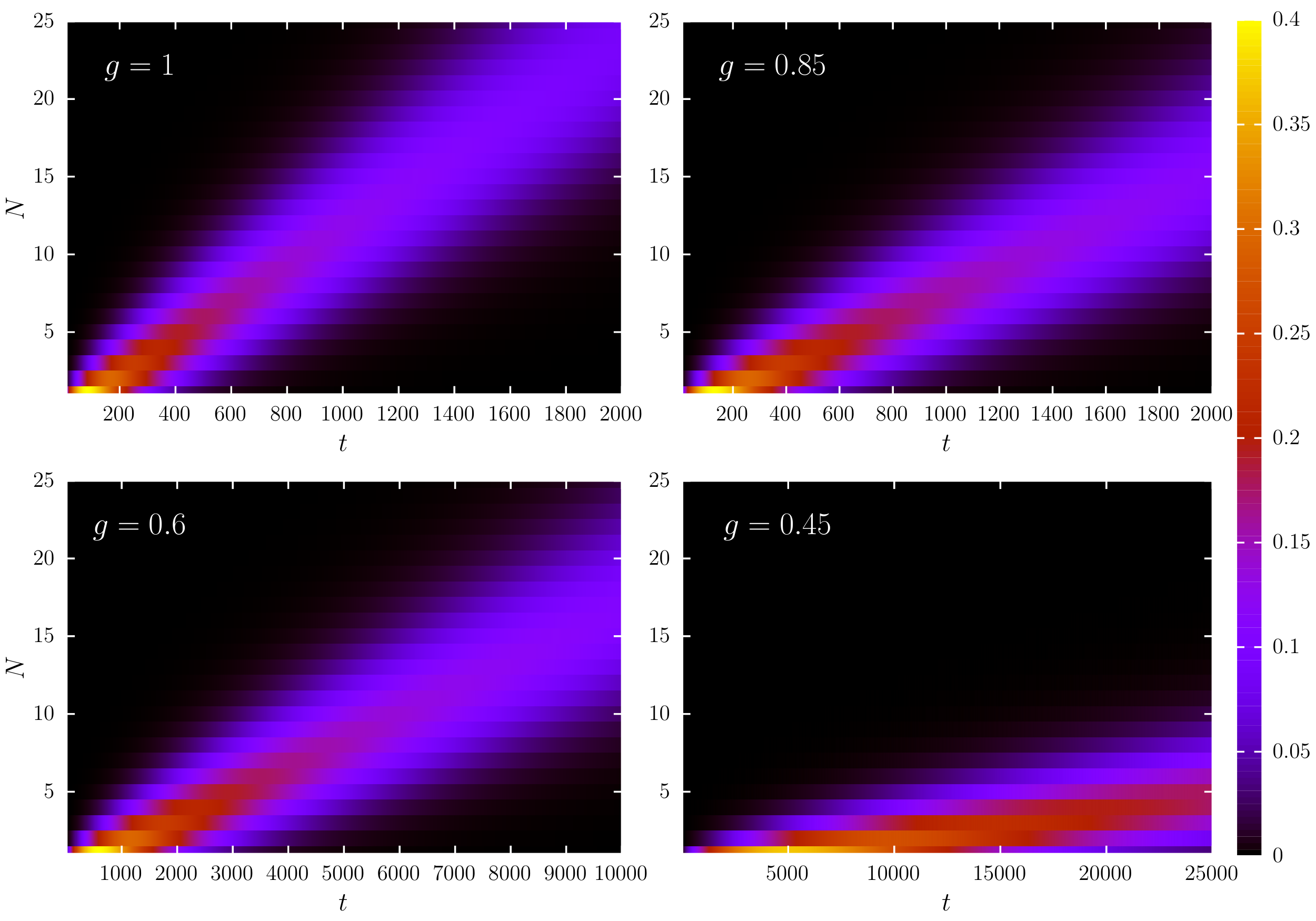}
\caption{Tunneling maps: the intensity plot of the distribution function $P(N,t)$ from Eq.(\ref{eq11}) in the low-temperature regime $ T\ll eV $ for the four different values of Luttinger liquid correlation parameter $g$: (non-interacting and weakly interacting electrons corresponds to $g=1$ and $g=0.85$ (left- and right-up maps, while the strongly interacting case corresponds to $g=0.6$ and $g=0.45$ (left- and right-bottom maps). Here for all four maps we set: $T=0.001 \Lambda$; $eV=0.5 \Lambda$. (In all calculations we set $\tilde{\lambda}^{2}=0.025 \Lambda^{2}$ while all energies are in the units of $ \Lambda \simeq E_{F}$ and all time scales - in the units of $ \delta = \hbar/\Lambda $ - the shortest time scale of the model.)}
\label{fig1}
\end{figure*}

Hence, the main goal of our study is to examine probability distribution function $P(N,t)$ defined as the inverse Fourier transform of generating functional over counting field $\xi$ \cite{19}
\begin{equation}
P(N,t)=\frac{1}{2\pi}\int\limits_{-\pi}^{\pi}d\xi e^{\mathcal{F}(\xi ,t)-\text{i}N\xi}.
\label{eq6}
\end{equation}  
As one can see from the results of Ref.~\cite{20}, the exact $\xi$- and $ t $- dependence of $\mathcal{F}(\xi,t)$ is highly non-trivial. However, in Ref. \cite{20} it has been noticed that following equality 
\begin{equation}
\mathcal{F}(\xi ,t)=t\cdot W_{B}(t).
\label{eq7}
\end{equation}
maintains even beyond the most commonly used steady state regime \cite{6} of electron transport at $ t \rightarrow \infty $. In particular, it has been justified in Ref.~\cite{20} that (\ref{eq7}) is valid for arbitrary $ t > 1/\Lambda_{g} $. As well in Ref.~\cite{20} it has been derived that function $W_{B}(t)$ in Eq.~(\ref{eq7}) has following \textit{exact} form \cite{20} 
\begin{equation}
\begin{aligned}
&W_{B}(t)=\Gamma_{g,+}(t)[e^{\text{i}\xi}-1]+\Gamma_{g,-}(t)[e^{-\text{i}\xi}-1]\\
&\Gamma_{g,\pm}(t)=\frac{\tilde{\lambda}^2}{2}\left( \frac{\pi T}{\Lambda_{g}}\right)^{2/g}\left[ J_{C}(t)\pm J_{S}(t)\right] ,\\
\end{aligned} \label{eq8},
\end{equation}
where time-dependent functions $\Gamma_{g,\pm}(t)$ are linear combinations of integrals $J_{C(S)}$ 
\begin{equation}
\begin{aligned}
&J_{C}(t)=\int\limits_{0}^{t}ds\frac{\cos(eVs)\cos(\pi/g-2\eta(s)/g)}{\big[\sinh^{2}(\pi Ts)+(\pi T/\Lambda_{g})^{2}\cosh^{2}(\pi Ts)\big]^{1/g}}\\
&J_{S}(t)=\int\limits_{0}^{t}ds\frac{\sin(eVs)\sin(\pi/g-2\eta(s)/g)}{\big[\sinh^{2}(\pi Ts)+(\pi T/\Lambda_{g})^{2}\cosh^{2}(\pi Ts)\big]^{1/g}},
\end{aligned}
\label{eq9}
\end{equation}
here $\eta(t)=\arctan\big[(\pi T/\Lambda_{g})\coth(\pi T t) \big]$ (see also the Appendix C in Ref.~\cite{20}) (it is worth noting that the case of $\Gamma_{g,-}=0$ corresponds to $W_{B}$ being characteristic function of Poissonian distribution). Considering (\ref{eq6}-\ref{eq8}) we arrive at
\begin{eqnarray}
P(N,t)=e^{-t\Gamma_{s}(t)}\sum_{k=0}^{\infty}\sum_{l=0}^{\infty}
\frac{t^{(k+l)}\left[ \Gamma_{g,+}(t)\right]^{k}\left[\Gamma_{g,-}(t)\right]^{l}}{k!l!} \nonumber \\
\times \frac{1}{2\pi}\int\limits_{-\pi}^{\pi}d\xi e^{\text{i}(k-l-N)\xi}. \ \ \ \ \ \ \
\label{eq10}
\end{eqnarray}
where we have introduced function $\Gamma_{s}(t)=\Gamma_{g,+}(t)+\Gamma_{g,-}(t)$. 
After that, time-dependent probability distribution function (\ref{eq10}) finally takes the following form
\begin{equation}
P(N,t)=e^{-t\Gamma_{s}(t)}\left[ t\Gamma_{g,+}(t)\right]^{N}\sum_{l=0}^{\infty}\frac{t^{2l}\left[ \Gamma_{g,+}(t)\Gamma_{g,-}(t)\right]^{l}}{l!(l+N)!}. \label{eq11}
\end{equation}
Since in (\ref{eq11}) functions $t\Gamma_{g,\pm}(t) \ll 1$ at any $ t $, the corresponding series is reasonably good convergent. Therefore, in our numerical calculations we take into account first 100 terms for the infinite sum in (\ref{eq11}) which provides a good precision.

On the four panels of Fig.~\ref{fig1} the  \textit{electron tunneling maps}, i.e. the 3D intensity graphs for the function $ P(N,t)  $ are plotted in the plane  $ (N; t)  $ in the low-temperature regime of electron transport through Luttinger liquid QPC at $T \ll eV$ for the four different values of Luttinger liquid correlation parameter (see Fig.~\ref{fig1} for details). For the four corresponding tunneling maps in three different temperature regimes, as well as for the additional details and general analysis of all distributions $ P(N,t) $ being obtained, one can see Appendix B in the Supplementary material. It is worth to note here that general behaviour of $ P(N,t) $ distributions from our tunneling maps is indirectly confirmed by the numerics obtained earlier in Ref.~\cite{21}.  

Then, from all panels of Fig.~\ref{fig1} (as well as from all panels of Figs.2a)-c) from the Appendix B below) one can see the main qualitative effect obtained in this paper: the presence of the regions of the \textit{anomalously high tunneling}(orange and red areas on tunneling maps Figs.1,2) on the short time scales beyond the steady state regime. These regions are direct consequences of the real-time \textit{self-equilibration} phenomenon in one-dimensional QPCs which was revealed for the first time in Ref.\cite{20}. It is worth to notice here that certain fingerprints of the same phenomenon of anomalous tunneling beyond the steady state were present in real time fluctuations of tunnel current in helical integer quantum Hall states depicted on Fig.4 from Ref.\cite{23}. All this confirms the universality of the self-equilibration and anomalous tunneling phenomena for the picture of electron tunneling in real time along the two tunnel connected Luttinger liquid channels of different nature. Surprisingly, from the two bottom panels of Fig.1 one can see that at $ eV \gg T $ the spreads of the (orange and red) anomalous tunneling regions beyond the steady state crucially depend on the small differences in $ g $  values if parameter $ g $ is between $ 0.6 $ and $ 0.4 $ (strong electron-electron interaction in the leads), while these spreads practically do not depend on small $ g $ deflections when $ 0.8 \lesssim g \leq 1 $. In the former (strongly interacting ) case  (for $ 0< g \lesssim 0.6 $) since the \textit{self-equilibration} phenomenon \cite{20} guarantees well-defined average current in the form (see e.g. Refs.~\cite{9,16,20})  $ \bar{I}_{g}\propto \left( eV/ \Lambda_{g}\right)^{(2/g - 1)}  $ at low temperatures even on short-time scales beyond the steady state regime one can define in this limit any probability $  P_{a,g} $  from the region of anomalous tunneling at certain fixed value of $ g $ as follows: $ P_{a,g}=\bar{I}_{g}\Delta\tau_{a} $, where $\Delta \tau_{a} $ is the characteristic "width" of the anomalous tunneling region along $ t $-axis on the corresponding tunneling map. Then for two different values of $ g $, say, $ g_{1}=g $ and $ g_{2} $ (where $ g_{2}=g_{1}+\Delta g $, $ \Delta g \ll g_{1}=g $ ) one can write down from corresponding tunneling maps that $ P_{a1,g}=P_{a2,g+\Delta g} $, which results in a following remarkable proportion
\begin{equation}
\frac{\Delta\tau_{a1}}{\Delta\tau_{a2}}\approx \left( \frac{eVg}{\Lambda}\right) ^{\frac{2}{g}-\frac{2}{(g+\Delta g)}}.
\label{eq12}
\end{equation}
Or, introducing a new parameter $ \sigma_{a}=\frac{\Delta\tau_{a1}}{\Delta\tau_{a2}} \leq 1$ under assumption $ \Delta g \ll g $, one obtains following non-trivial estimation for $  \Delta g $ dependent on measured $ \sigma_{a} $
\begin{equation}
\Delta g \approx \frac{g^{2}}{2} \left \vert \frac{\ln(\sigma_{a})}{\ln\left( \frac{eVg}{\Lambda}\right)}\right \vert.
\label{eq13}
\end{equation}
Obviously, for two bottom panels of Fig.~\ref{fig1} (with $ g_{2}=0.6 $ and $ g_{1}=g=0.45 $) the value of $ \sigma_{a} $ is approximately $ 0.1 $, while  $ eV/\Lambda=0.5 $, hence, it is easy to see that estimation (\ref{eq13}) gives correct prediction of $ \Delta g=g_{2}-g_{1} $ for these two maps. In particular, from Eq.~(\ref{eq13}) with excellent precision of $ \pm 0.01$ it follows for two bottom panels of Fig.~\ref{fig1} that $ \Delta g \approx 0.15$ as it should be. 

To conclude, in the above we have derived and plotted as 3D tunneling maps the distribution functions $ P(N,t) $ for $ N $ electrons to tunnel during the fixed time interval $ t $ through arbitrary weakly linked quantum-point contact (QPC) with arbitrary Luttinger liquid 1D leads in a whole range of bias voltages and temperature values. This derivation is based on the Self-equilibration theorem being proven in Ref.~\cite{20}.
The main qualitative effect being predicted is the presence of "tunneling anomalies" - regions with high tunneling probabilities for small numbers $ N $ of electrons tunnelled through the tunnel junction on short-time scales. 
It is shown for the first time that these tunneling anomalies allow for a precise estimation of small differences in values of Luttinger liquid correlation parameter $ g $ for 1D quantum-point contacts those characteristics are close to each other.
This study is a part of research project "Novel aspects of quantum detectors theory in low-dimensional electron systems" funded by the research grant No.09/01-2020 and 09/01-2021(2) from the National Academy of Sciences of Ukraine. 


\section{Supplementary material:}

\subsection*{Appendix A: Bosonic field operators in the interaction representation}
\label{App_A}
\renewcommand{\theequation}{A.\arabic{equation}}
\setcounter{equation}{0}

Within the bosonization language \cite{9,10,16} quantum bosonic fields $ \varphi_{L(R)}(x) $ and $ \theta_{L(R)}(x) $  correspond to fluctuating parts of current- and charge electron densities in the QPC as $ j_{L(R)}(x)=\partial_{x}\varphi_{L(R)}(x)=\sum_{c=\pm}c:\Psi_{c,L(R)}^{\dagger}(x)\Psi_{c,L(R)}(x): $ and  $ \rho_{L(R)}(x)=\partial_{x}\theta_{L(R)}(x)=\sum_{c=\pm}:\Psi_{c,L(R)}^{\dagger}(x)\Psi_{c,L(R)}(x): $ correspondingly. Here $c=\pm$ is a chirality of left- (or right-) moving electron and symbol $:(\dots):$ corresponds to a normal ordering of standard fermionic creation (annihilation) field operators
$\Psi_{c,L(R)}^{\dagger}(x)$\big($\Psi_{c,L(R)}(x)$\big). Bosonization relation between fermionic field operators $\Psi_{c,j}(x,t)$ and bosonic quantum fields $ \theta_{j}(x)$ and $\varphi_{j}(x) $ ($ j=L,R $ and $ c=\pm $) is standard \cite{6,9,11,20} 
\begin{equation}
\Psi_{c,j}(x,t)=\frac{\eta_{c,j}}{\sqrt{2\pi a_{0}}}e^{\text{i}c(k_{F}-\pi/L)x}e^{\mp\text{i}eVt/2}e^{\text{i}(c\theta_{j}(x)+\varphi_{j}(x))}
\label{eqA1}
\end{equation}
In Eq.~(\ref{eqA1}) the length parameter $ a_{0} $ is of the order of the lattice constant in QPC electrodes, this is the shortest length scale of the model. 

In the fermionic representation the tunnel Hamiltonian of arbitrary weakly linked QPC reads\cite{6}  
\begin{equation}
H_{\text{int}}=t_{t}:\Psi_{+,R}^{\dagger}\Psi_{+,R}(0,t): +~ t_{t}^{*}:\Psi_{-,L}^{\dagger}\Psi_{-,L}(0,t):,
\label{eqA2}
\end{equation}
where $t_{t}$ and $t_{t}^{*}$ are bare tunneling amplitudes. Making use of the bosonization formula (\ref{eqA1}) and boundary condition (\ref{eq3}) one can easily bosonize the fermionic tunnel Hamiltonian (\ref{eqA2}) in terms of spatially non-local bosonic fields $ \varphi_{\pm}(x) = \varphi_{L}(x) \pm \varphi_{R}(x)$ to its form of Eq.(\ref{eq4}) from the main text. 

In what follows we will consider our model (\ref{eq1}) within the interaction representation with respect to the bosonized interaction Hamiltonian (\ref{eq4}). In this picture bosonic phase fields become explicitly time dependent operator functions $ \varphi_{\pm}(x,t) $ where the corresponding time dependence is dictated by the free Luttinger liquid Hamiltonian (\ref{eq2}). Secondary quantization of bosonic spatially non-local quantum fields $ \varphi_{\pm}(x,t) $  yields \cite{6,11,20} 
\begin{equation}
\varphi_{\pm}(x,t)=\sum_{k\neq 0} \frac{\text{i}\sqrt{\pi}\cdot e^{-\vert k \vert a_{0}/2}}{\sqrt{2gL \vert k \vert}} \left[ e^{-\text{i}kx}b^{+}_{k,\pm}(t) - e^{\text{i}kx}b_{k,\pm}(t) \right] 
\label{eqA3}
\end{equation}
where secondary quantized  bosonic operators $b^{+}_{k,\pm}(t)$ (and $b_{k,\pm}(t) $)   create (and annihilate) plasmonic modes with momenta $ k $ and each plasmonic mode is delocalized along the entire infinite one-dimensional system.  Since in the secondary quantized representation (\ref{eqA3}) the Luttinger liquid Hamiltonian (\ref{eq2}) of the leads takes canonical quadratic form for free bosons (or plasmons) \cite{6,11}
\begin{equation}
H_{\text{LL}}=\sum_{\alpha=\pm,k=0}^{\infty}v_{g}|k|b_{k,\alpha}^{\dagger}(0)b_{k,\alpha}(0)
\label{eqA4}
\end{equation}
the solutions of corresponding Heisenberg quantum equations of motion (QEOMS) for operators $b^{+}_{k,\pm}(t)$ (and $b_{k,\pm}(t) $) with free bosonic Hamiltonian (\ref{eqA4}) describe creation (annihilation) of a free delocalized plasmons with momentum $ k $ and linear dispersion, propagating along the infinite 1D system in both directions \cite{11}  ($ c=\pm $) $b^{+}_{k,\pm}(t)=b^{+}_{k,\pm}(0)e^{\text{i}v_{g}t\vert k \vert}$ (and $b_{k,\pm}(t)=b_{k,\pm}(0)e^{-\text{i}v_{g}t\vert k \vert} $). Bosonic operators  $b^{+}_{k,\pm}(0)$ (and $b_{k,\pm}(0) $) for the initial moment of time $ t=0 $ fulfil canonical bosonic commutation relations $ [ b_{k,\alpha}(0), b^{+}_{k',\alpha'}(0)] = \delta_{k,k'}\delta_{\alpha,\alpha'} $ (where $ \alpha,\alpha' = \pm $) and give standard thermally averaged vacuum expectation values  $ \langle b^{\dagger}_{k,\alpha}(0)b_{k^{\prime},\alpha^{\prime}}(0)\rangle = n_{b}(k)\delta_{k,k^{\prime}}\delta_{\alpha,\alpha^{\prime}} $ where $n_{b}=(e^{v_{g}|k|/T}-1)^{-1}$ is Bose-Einstein distribution function for free plasmons at arbitrary temperature $ T $ of the QPC environment. Substituting these free operator solutions into Eq.~(\ref{eqA3}) and then substituting (\ref{eqA3}) into Eq.(\ref{eq4}) of the main text one obtains basic QPC interaction Hamiltonian $ H_{\text{int}}(t) $ being explicitly time-dependent strongly non-linear operator function in the interaction representation \cite{20}.

\subsection*{Appendix B: Tunneling maps of interacting electrons in different temperature regimes}
\renewcommand{\theequation}{B.\arabic{equation}}
\setcounter{equation}{0}

On the Figs.~\ref{fig2},a)-c) the probability distribution function $ P(N,t) $ from Eq.(\ref{eq11}) in  three different regimes is plotted: i) for the high-temperature limit: $T \gg eV$ (see Fig.~\ref{fig2},a), ii) for the intermediate temperatures: $T \simeq eV$ (see Fig.~\ref{fig2},b) and iii) for the low-temperature limit: $T \ll eV$ (see Fig.~\ref{fig2},c).  All those three scenarios are represented on the Figs.~\ref{fig2}, a)-c) in the form of corresponding \textit{electron tunneling maps} plotted in the plane  $ (N;t) $  for different values of parameter $ eV/T $ (Fig.~\ref{fig2},a)-c)) and - for each a) to c) case - for the four different values of Luttinger liquid correlation parameter $ g $ in the QPC electrodes. The latter sub-cases for each a) to c) case are: non-interacting electrons ($g=1$, left-up map), weakly interacting electrons ($g=0.85$, right-up map) and strongly interacting electrons ($g=0.6$, left-bottom map) and $g=0.45$ ( right-bottom map). On the Fig.~\ref{fig2}, d) the distributions $ P(N) $ which are the cross-sections of all tunneling maps from Fig.~\ref{fig2},c) by the line $ t=1000\delta $ for different values $ g $ are shown. Obviously, from all the data plotted on Fig.~\ref{fig2} one can find also the maximum $ P(N_{\text{max}},t) $ of the probability distribution function $ P(N,t) $ which is achieved at certain "optimal" value $ N_{\text{max}} $ of electrons have been tunneled by a chosen moment of time $t> \delta_{g}=1/\Lambda_{g}$  for the four different values of $ g $ . Then analysing the $N$-dependent term in (\ref{eq11}) $[t\Gamma_{g,+}(t)]^{N}/(N)!$ one can get the following relation between $N_{\text{max}}$ and $t$   
\begin{equation}
\ln(t \Gamma_{g,+}(t))=\psi(N_{\text{max}}+1),
\label{eqB1}
\end{equation}
where $\psi(x)$ is a well-tabulated digamma function. The appearance of a function $\psi(z)$ in the average tunnel current calculations is well-known in the literature (see, e.g. Refs. \cite{9,16,17}), it signals about the correctness of all the tunnel probability maps calculations represented on Figs.~\ref{fig1},~\ref{fig2}, a)-c). The solutions of Eq.~(\ref{eqB1}) are plotted on Fig.~\ref{fig1},e) for different values of $ g $. Remarkably, the constant ratios $ N_{\text{max}}(t)/t $ which are depicted on Fig.~\ref{fig2},e) as straight lines in the $ {N;t} $ plane are the evidences of  well-defined average currents \cite{9,10,16,17} on the entire time scale of observations $ t > 1/\Lambda_{g} $ including the short-time scale which is beyond the steady state regime of electron tunneling. The latter fact is a direct consequence of the Self-equilibration phenomenon being predicted in Ref. \cite{20} for such type of quantum systems.

\begin{figure*}
\includegraphics[width=\columnwidth]{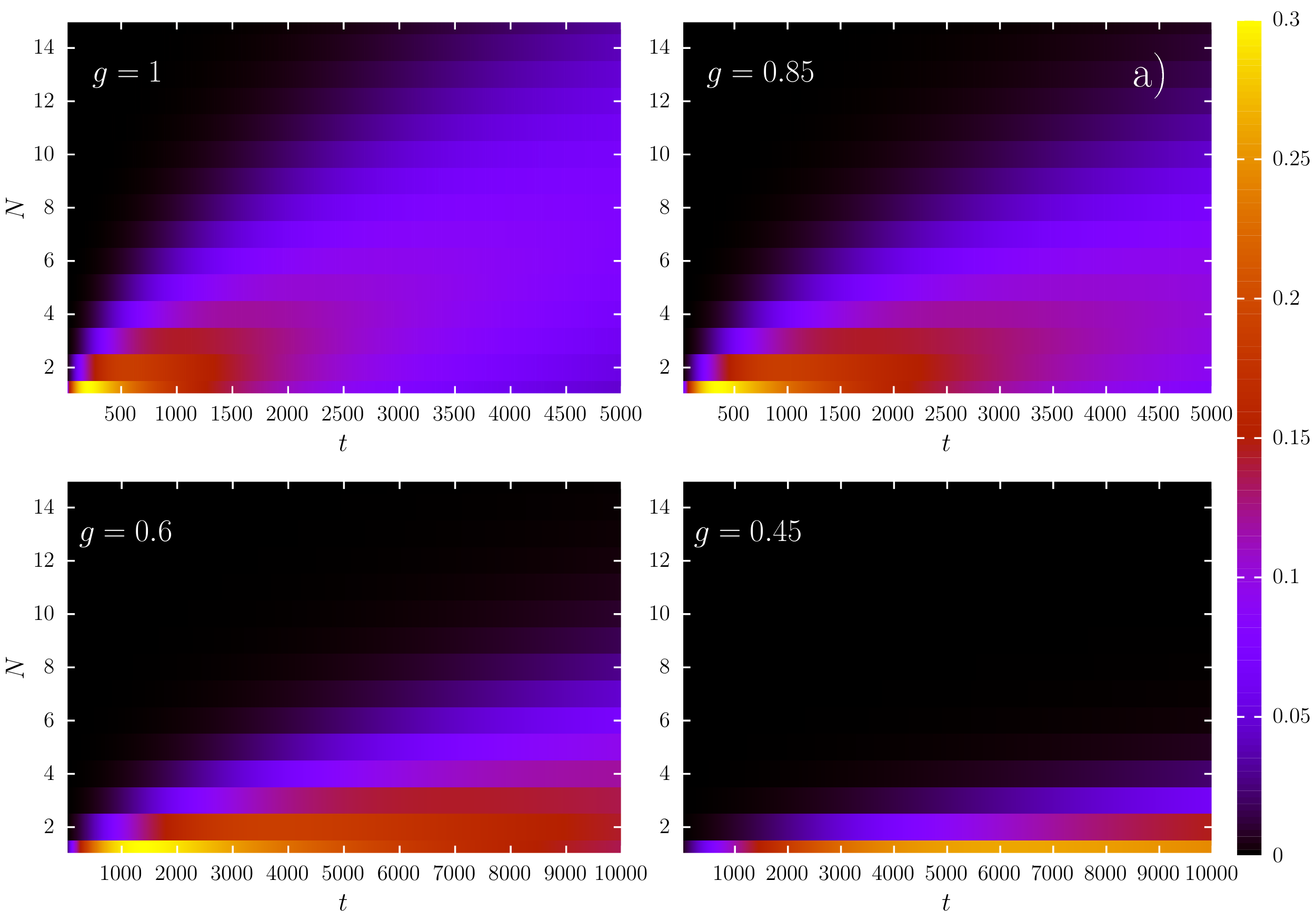}
\includegraphics[width=\columnwidth]{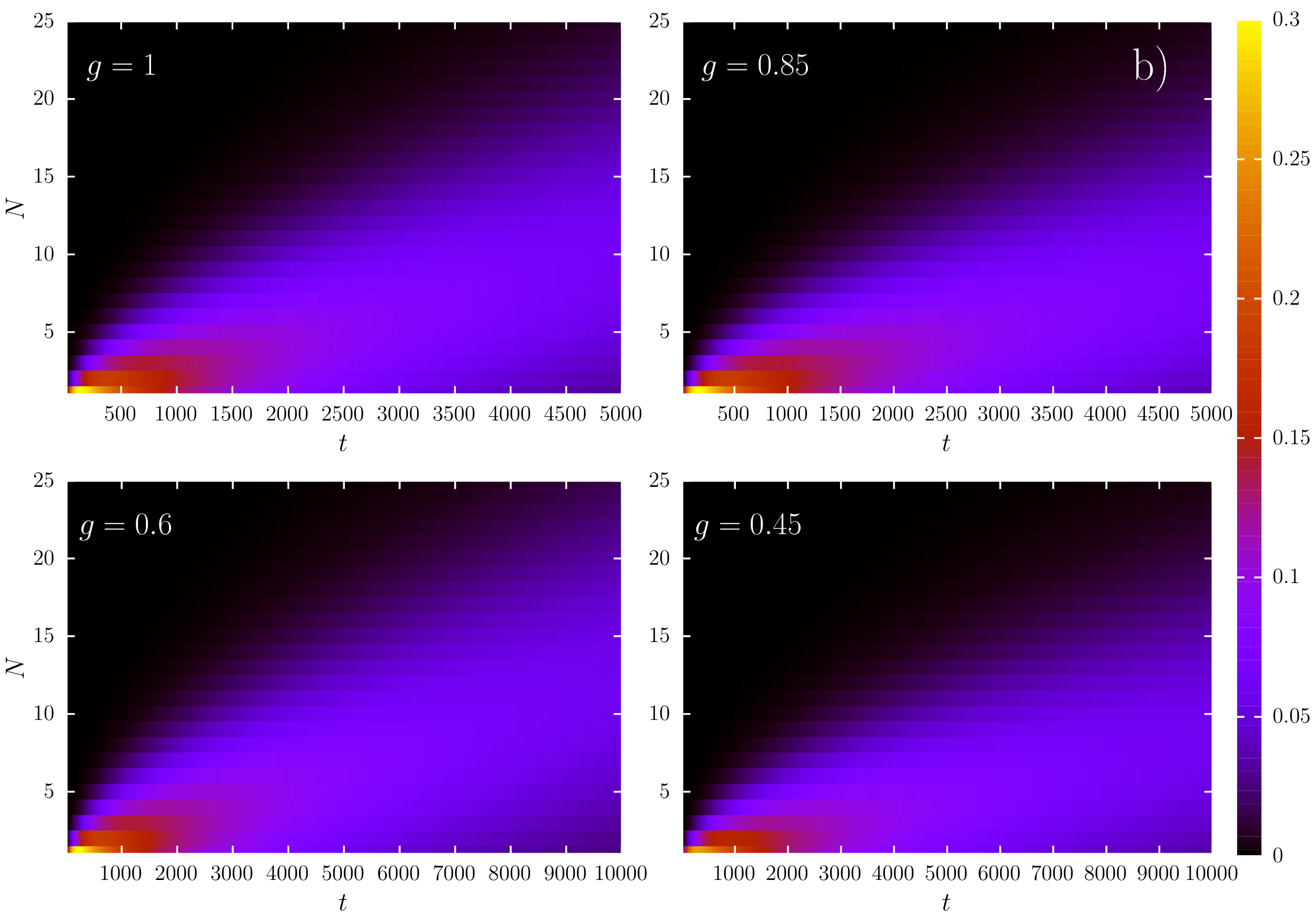}
\includegraphics[width=\columnwidth]{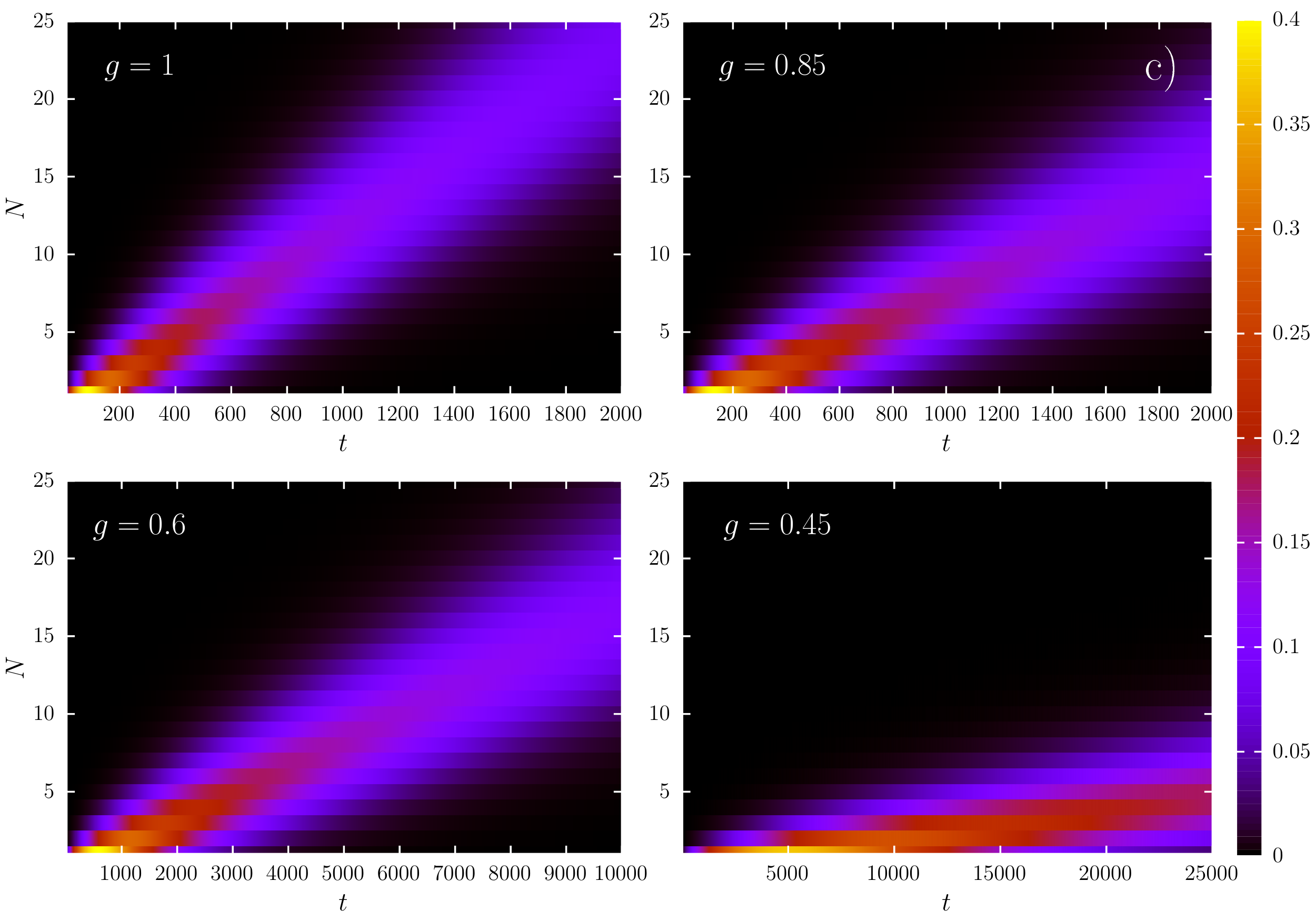}
\includegraphics[width=\columnwidth]{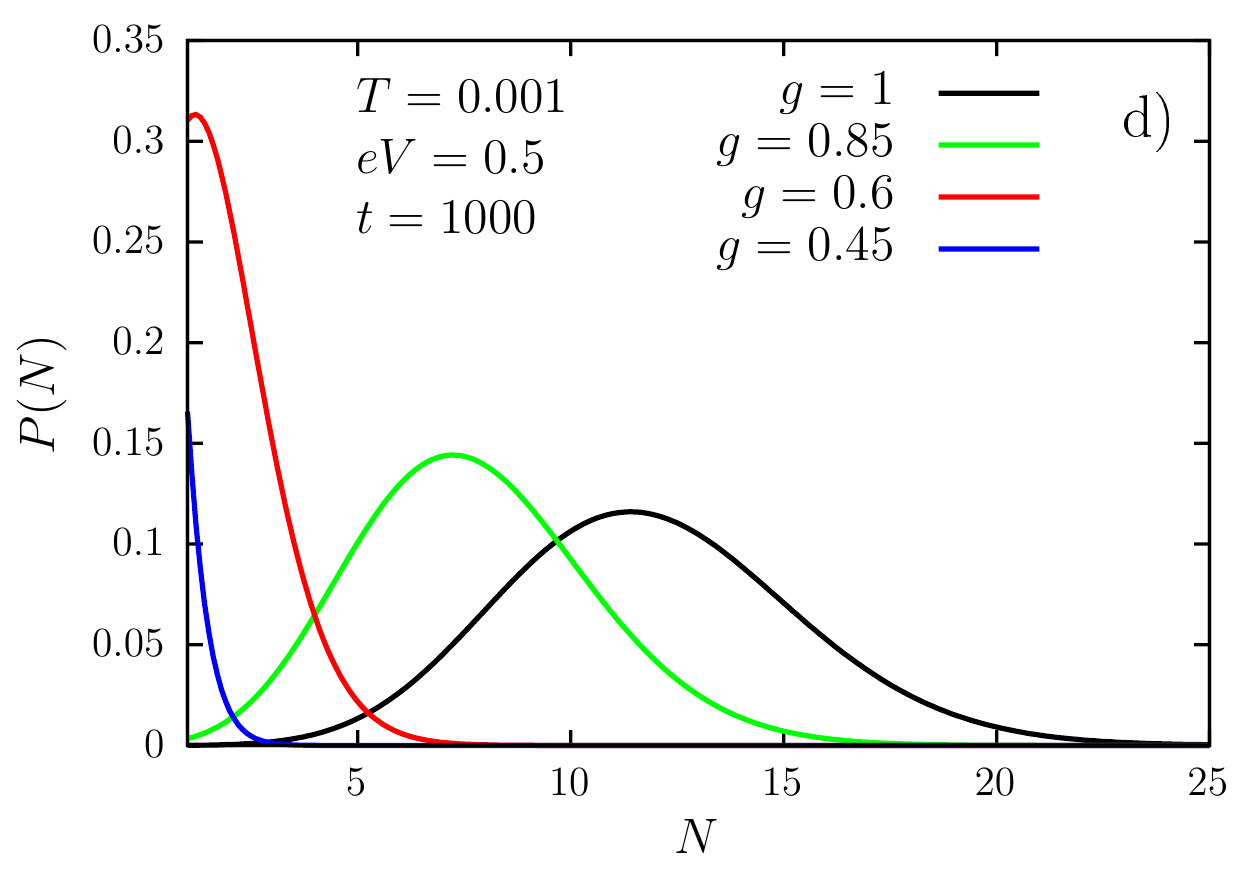}
\includegraphics[width=\columnwidth]{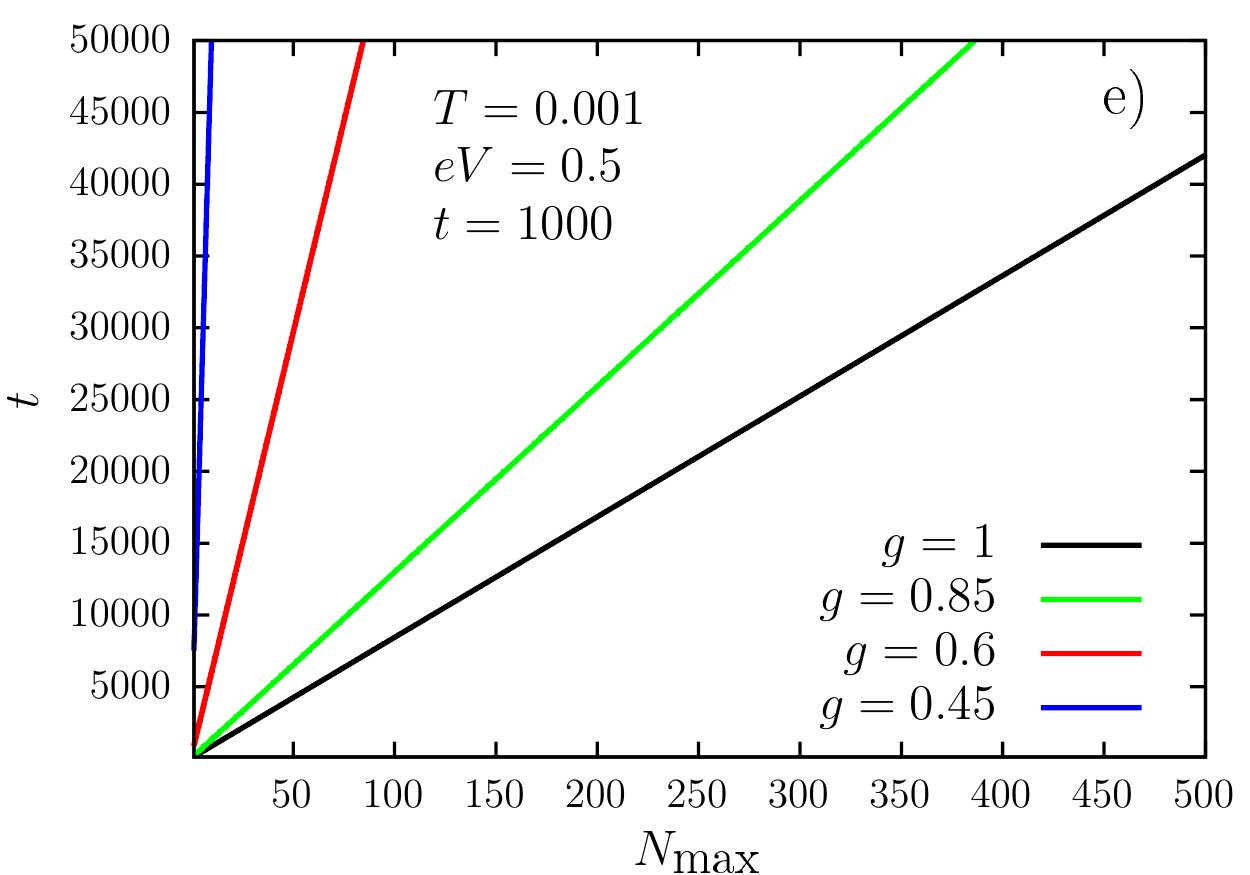}
\caption{a)-c) Tunneling maps: the intensity plot of the distribution function $ P(N,t) $ from Eq.(\ref{eq11}) for different values of Luttinger liquid correlation parameter $g$ (non-interacting and weakly interacting electrons corresponds to $g=1$ and $g=0.85$ (left- and right-up maps for each a) to c) case, while the strongly interacting case corresponds to $g=0.6$ and $g=0.45$ (left- and right-bottom maps for each a) to c) case). Here for all a) maps one has $T=0.1\Lambda$; $eV=0.05\Lambda$  (left up panel), for all b) maps: $T=0.5\Lambda$; $eV=0.5\Lambda$ (right up panel) and for all c) maps: $T=0.001\Lambda$; $eV=0.5\Lambda$ (left-bottom panel). (In all calculations we set $\tilde{\lambda}^{2}=0.025\Lambda^{2}$ while all energies are in the units of $ \Lambda \simeq E_{F}$ , where $ E_{F} $ is the Fermi-energy of the QPC leads and all time scales are in the units of $ \delta = \hbar/\Lambda $ -the shortest time scale in the model. d) Cross-section of tunneling maps from Fig.~\ref{fig2} c) by the line $t=1000\delta$ for different values of $g$ (right bottom panel). e) Time dependence of $N_{\text{max}}$ which corresponds to the maximum of probability distribution function $ P(N,t) $ from Eq.(\ref{eq11}). Corresponding data obtained by solving transcendental equation $\ln(t\Gamma_{g,+}(t))=\psi(N_{\text{max}}+1)$, where $\psi(z)$ is digamma function.}
\label{fig2}
\end{figure*}

More generally, from all real-time tunneling maps being plotted on Figs.~\ref{fig1},~\ref{fig2}, a)-c) one immediately sees the most common features of tunnel electron transport in one-dimensional quantum-point contacts at different temperatures and different electron-electron interaction in the QPC leads. For example, one sees that as the result of a gradual temperature decrease (see Figs.~\ref{fig2}, a)-c)) non-vanishing tunneling probabilities become concentrated in the narrow corridor (light blue area on all maps) around the straight line in the $ (N;t) $ plane of the maximal tunneling probability, which marks the constant average current (i.e. the ratio $ N_{\text{max}}/t_{\text{max}}=\bar{I}=const $ ), while the rise of the temperature smears the region of non-zero electron tunneling to the larger values of $ t $ for the same fixed $ N $ chosen. At the same time the entire region of non-vanishing tunneling probabilities (i.e. the entire blue area on tunneling maps) shrinks while one heats the system. These effects are because the increase of thermal fluctuations suppresses average tunnel current spreading its dispersion. The appearance of the region of the highest tunneling probability (orange and red areas) on all tunneling maps for relatively small values of $ N $ (for $1 \leq N \lesssim 8 $) and $ t $ (for $  \Delta t \lesssim 10^{2}\delta$ or $  \Delta t \lesssim 10^{4}\delta$ dependent on the interaction and temperatures in the system) - is nothing else but a remarkable evidence of a new regime of \textit{anomalous electron tunneling} on short-time scales beyond the steady state regime in \textit{all types} of one-dimensional quantum-point contacts. This regime should be referred to as the effect of strong electron correlations of special type \cite{11,20}  in one-dimensional system which should drive the system to its steady state regime in the process of its \textit{self-equilibration}(see Ref.~\cite{20} on further details of the self-equilibration phenomenon). 

At the same time the manifestation of the effects of electron-electron interaction (i.e. the effects one observes varying $ g $ between $ g=1 $ (non-interacting electrons) and $ 0< g \ll 1 $ (strongly repulsing electrons in the leads of QPC) on all obtained tunneling maps turns out to be even more bright. First, as one can see from all maps on Fig.~\ref{fig1} the decrease of $ g $ diminishes the average current, i.e. the angle between the "direction" of the orange-blue "flame" on the maps on Fig.~\ref{fig1} and the $ (0;x) $ axis. This is nothing more than a famous  Kane-Fisher effect of tunnel current suppression \cite{10,16} in strongly interacting Luttinger liquid tunnel contacts at low temperatures. 

As well, the most interesting effect of strong electron-electron interaction ($ g \lesssim 1/2 $) in the leads of Luttinger liquid QPCs is evident from the analysis of tunneling maps on the two bottom panels of Fig.~\ref{fig1} as well as from two bottom maps on Fig.~\ref{fig2},c). In the strongly interacting case (for $ 0< g \lesssim 0.6 $) the small difference $ \simeq 0.1 $ in $ g $ values results in a huge differences in the anomalous tunneling region widths, which can be of the order of $ 9000\delta $ (!). The latter effect is a brand novel feature of the interaction -dependent anomalous tunneling for the small numbers of electrons and it can be used for a precise estimations of small differences in values of $ g $ interaction parameter for different one-dimensional QPCs from the tunneling experiments on short-time scales by comparing the experimental tunneling maps with ones from Figs.~\ref{fig1},~\ref{fig2} (see also the main text for related explanations).


\end{normalsize}

\begin{references}
\bibitem{22}
Here it is worth to notice that, generally speaking, the bare tunnel amplitudes $ t_{t} $ renormalization by means of counting field $ \xi $ can be made in any form of the kind  $ t_{t} \rightarrow t_{t} \exp(-\text{i} \xi /m) $ with any positive integer $ m=1,2,3,.. $. In our (most common) case we have $ m=1 $, but also for any $ m > 1 $ the corresponding renormalization leads to the same distributions $ P(N,t) $ if one simultaneously renormalizes $ N \rightarrow m N $ and $ \delta \rightarrow m \delta $. The latter property is due to the scalability of the entire FCS scheme to larger electron numbers and larger time scales. For example, in Ref.~\cite{20} it is presumed that $ m=2 $ without any effect on corresponding correlators being calculated there.
\end{references}
\end{document}